\documentclass{jltp}
\usepackage{amssymb,amsmath}

\title{Hall effect in quasi one-dimensional organic conductors}

\author{Gladys Le\'{o}n and Thierry Giamarchi}

\address{University of Geneva, 24 Quai Ernest-Ansermet CH-1211 Geneva 4, Switzerland}

\runninghead{G. Le\'{o}n and T. Giamarchi}{Hall effect in quasi one-dimensional organic conductors}

\begin{document}

\maketitle

\begin{abstract}
We study the Hall effect in a system of weakly coupled Luttinger
Liquid chains, using a Memory function approach to compute the Hall
constant in the presence of umklapp scattering along the chains. In
this approximation, the Hall constant decomposes into two terms: a
high-frequency term and a Memory function term. For the case of zero
umklapp scattering, where the Memory function vanishes, the Hall
constant is simply the band value, in agreement with former results
in a similar model with no dissipation along the chains. With
umklapp scattering along the chains, we find a power-law temperature
dependance of the Hall constant. We discuss the applications to
quasi 1D organic conductors at high temperatures.
\end{abstract}

\section{INTRODUCTION}

Within the organic conductors, the Bechgaard salts $(TMTSF)_{2}X$,
have been widely studied for their many interesting properties and
their complex phase diagram.\cite{jerome_review_chemrev} In
particular their low dimensionality makes them very special: they
experience a dimensional crossover from a 1D Luttinger
Liquid\cite{schwartz_electrodynamics} (LL) to a higher dimensional
metallic state\cite{giamarchi_review_chemrev}. An important open
question is the Hall effect in such an interacting system made of
coupled chains. At low temperature theories based on a Fermi liquid
description are possible\cite{yakovenko_phenomenological_model_exp}.
In the high temperature coupled LL regime a description of the Hall
effect was only possible in the quite restricted case where no
scattering exists inside the
chains\cite{lopatin_q1d_magnetooptical}. In that case one
surprisingly obtains a temperature independent Hall resistance given
by the simple band value even if interactions are present in the
system. Up to now, no theory when scattering is present in the
chains was available.
Measurements\cite{moser_hall_1d,mihaly_hall_1d} were made in the
compound $(TMTSF)_{2}PF_{6}$ giving different results for the Hall
constant above the dimensional crossover temperature and making the
need for a theoretical description even more important. We present
here such a theory of the Hall effect of coupled LL. We use a Memory
matrix approach to compute the Hall constant and find out its
temperature dependance.

\section{MODEL}

We consider a model of coupled LL chains (along the x-axis) with
umklapp scattering, coupled by a transverse hopping (y-axis) and a
magnetic field applied perpendicular to the plane of the chains
(z-axis). The hamiltonian $\hat{H}=\hat{H}_{0}+\hat{H}_{1}$ for
fermions with spins, decomposes into a unperturbed part
$\hat{H}_{0}$ and a perturbation term $\hat{H}_{1}$. The first part
is:
\begin{multline}\label{Hamiltonian}
 \hat{H}_{0}=\int dx\left[\sum_{i,\sigma}v_{F}\hat{\psi}_{i,\sigma}^{\dagger}
 \tau_{3}(-i\partial_{x})\hat{\psi}_{i,\sigma}-\alpha\sum_{i,\sigma}\hat{\psi}_{i,\sigma}^{\dagger}
 \partial_{x}^{2}\hat{\psi}_{i,\sigma}\right.\\
 \left.+U\sum_{i,\sigma,\sigma'}\rho_{i,\sigma}\rho_{i,\sigma'}-t_{\bot}\sum_{\langle i,j \rangle,\sigma}
 \hat{\psi}_{i,\sigma}^{\dagger}\hat{\psi}_{j,\sigma}e^{-i\frac{e}{c}A_{i,j}}\right],
\end{multline}
where $\hat{\psi}$ is a two component vector composed from the right
and left-moving electrons, $\tau_{3}$ is a pauli matrix and
$A_{i,j}=\int_{i}^{j}{\bold{A}}d{\bold{l}}$, where we use the Landau
gauge $A_{y}=Hx$. The second term in (\ref{Hamiltonian}) is a
nonlinear correction to the spectrum. Including such a band
curvature term is necessary due to the fact that in chains with
linear spectrum, the particle-hole symmetry gives a zero Hall
resistivity. The spectrum is thus $\epsilon_{\pm}=\pm
v_{F}\left(p\mp p_{F}\right)+\alpha\left(p\mp p_{F}\right)^{2}$. The
third term in (\ref{Hamiltonian}) are the momentum conserving
interaction processes\cite{giamarchi_book_1d}. The last term is the
coupling between the chains.

The interaction term $\hat{H}_{1}$ contains the {\it umklapp}
processes, where two particles are scattered from one side to the
other side of the Fermi surface, transferring or absorbing momentum
from the lattice. For simplicity we consider here the case of
half-filled commensurate system which satisfy
$4k_{F}=\frac{2\pi}{a}$, where $a$ is the lattice spacing along the
chains. Then $\hat{H}_{1}$ is
\begin{equation}
 \hat{H}_{1}=g_{3}\int dx\sum_{i,\sigma}\left(\hat{\psi}_{R,i,\sigma}^{\dagger}\hat{\psi}_{R,i,-\sigma}^{\dagger}
 \hat{\psi}_{L,i,-\sigma}\hat{\psi}_{L,i,\sigma}+ \mathrm{h.c.} \right)
\end{equation}
To compute the Hall constant $R_{H}$ we use a Memory function
approach, which consists on a perturbative expansion of the
resistivities\cite{mori_theory_book}. Within this approach an
expression of $R_H$ was obtained\cite{lange_hall_constant} for the
isotropic Hubbard model. In the case of coupled chain the
corresponding expression correctly taking the anisotropy into
account becomes\cite{leon_hall_short}
\begin{equation}\label{HallConstant}
 R_{H}(z)=\frac{1}{i\chi_{y}^{0}}\lim_{H\rightarrow0}\frac{\Omega_{xy}+iM_{xy}(z)}{H}
\end{equation}
where $\Omega_{xy}$, the frequency matrix, is defined as
$\equiv\frac{1}{\chi_{x}^{0}}\langle\left[\hat{J}_{x},\hat{J}_{y}\right]\rangle$.
This term represents the high-frequency limit and when there is no
perturbation (umklapp scattering) it gives back the exact result of
Ref~\onlinecite{lopatin_q1d_magnetooptical}. This check ensures that
the Memory function approximation indeed captures the correct
physics. The $\chi_{\nu}^{0}$ are the diamagnetic terms in the $x$
and $y$ directions. The second term in (\ref{HallConstant}) is the
Memory function, which is given by
\begin{equation}
 i\chi_{x}^{0}M(z)=-\frac{\langle\langle\hat{K}_x;\hat{K}_y\rangle\rangle_{z}^{0}}{z}
\end{equation}
The $\hat{K}_{\nu}$ operators are the commutator between the umklapp
term and the current along the $\nu$ direction
$\hat{K}_{\nu}=\left[\hat{H}_{1},\hat{J}_{\nu}\right]$.

\section{RESULTS}

The difficult part is to compute the correlation function
$\langle\langle\hat{K}_x;\hat{K}_y\rangle\rangle_{z}^{0}$. Technical
details will be presented elsewhere\cite{leon_hall_short} and we
quote here only the results. For weakly coupled $1/2$-filled LL
chains, the Hall constant has a power law dependance on temperature
\begin{equation}\label{HallConstant_result}
 R_{H}=R_{H}^{0}\left[1-A\left(\frac{g_{3}}{u_{\rho}}\right)^{2}
 \left(\frac{\alpha T}{u_{\rho}}\right)^{3K_{\rho}-3}\right],
\end{equation}
where $R_{H}^{0}$ is the band
value\cite{lopatin_q1d_magnetooptical}, $g_{3}$ is the umklapp
parameter, $u_{\rho}$ the velocity of the charge excitations along
the chains and $A$ a constant. The LL
parameter\cite{giamarchi_book_1d} $K_{\rho}$ controls the
temperature dependence (we take $K_{\sigma}=1$ for the spin part).
When there are no interactions $K_{\rho}=1$ giving no temperature
dependence for $R_{H}$. For repulsive interactions $K_{\rho} < 1$
(as in the organic conductors\cite{schwartz_electrodynamics}). In
this case two important limits appears: for small $g_{3}$ or large
temperature, $R_{H}\rightarrow R_{H}^{0}$. Note that the power law
dependence found for the Hall constant is
different from the the one found for the conductivity along LL
chains with umklapp scattering\cite{giamarchi_umklapp_1d} $\sigma(T)\propto
T^{3-4K_{\rho}}/g_{3}^{2}$.

\section{DISCUSSION AND CONCLUSIONS}

Using a Memory function approach, we obtained the Hall constant in
weakly coupled LL chains with umklapp scattering along the chains,
and we obtained a power law dependance for $1/2$-filled commensurate
systems. It would be interesting to compare such result with
measurements of the Hall resistance in purely half filled organic
compounds. To compare with the Bechgaard salts it is important to
take into account that these compounds have both half and quarter
filled commensurate character. If quarter filling umklapp
$g_{3,1/4}$ is present one expects the Hall resistance to be
\begin{equation}\label{HallConstant_2}
 \frac{R_{H}}{R_{H}^{0}}=1-A\left(g_{3,1/2}\right)^{2}T^{\mu}+B\left(g_{3,1/4}\right)^{2}T^{\nu}.
\end{equation}
where the calculation for $\nu$ will be presented
elsewhere.\cite{leon_hall_short} Longitudinal transport is
dominated\cite{schwartz_electrodynamics} by the quarter filled
commensurability and it will be interesting to check whether the
same is true for the Hall resistance. To decide on that point it
will however be useful to have additional measurements of Hall
resistance in order to reconcile the two different temperature
behaviors that are for now observed experimentally in the same
compound\cite{moser_hall_1d,mihaly_hall_1d}.

\section*{Acknowledgements}

This work has been supported in part by the Swiss National Fund for
research under MANEP and Division II.

\bibliographystyle{prsty}

\end{document}